\documentclass[fleqn,usenatbib]{mnras}

\usepackage{newtxtext,newtxmath}

\usepackage[T1]{fontenc}

\usepackage{graphicx}	
\usepackage{amsmath}	
\usepackage{amssymb}	

\newcommand{\abd}[1]{$\log \left(\textrm{#1}/\textrm{H}\right)$}
\newcommand \msun {M$_\odot$}

\title[WD variability and debris accretion]{Periodic optical variability and debris accretion in white dwarfs: a test for a causal connection\thanks{Based on observations made with the NASA/ESA \textit{Hubble Space Telescope}, obtained at the Space Telescope Science Institute, which is operated by the Association of Universities for Research in Astronomy, Inc., under NASA contract NAS 5-26555. These observations are associated with programmes 14082 (PI: Maoz), 11526 (PI: Green), and 14076 (PI: G\"{a}nsicke).}}

\author[N. Hallakoun et al.]{
Na'ama Hallakoun,$^{1,2}$\thanks{E-mail: \href{mailto:naama@wise.tau.ac.il}{naama@wise.tau.ac.il}}
Dan Maoz,$^{1}$
Eric Agol,$^{3}$
Warren R. Brown,$^{4}$
Patrick Dufour,$^{5}$
\newauthor{
Jay Farihi,$^{6}$
Boris T. G\"{a}nsicke,$^{7}$
Mukremin Kilic,$^{8}$
Alekzander Kosakowski,$^{8}$
}
\newauthor{
Abraham Loeb,$^{9}$
Tsevi Mazeh,$^{1}$
and Fergal Mullally$^{10}$
}
\\
\\
$^{1}$School of Physics and Astronomy, Tel-Aviv University, Tel-Aviv 6997801, Israel\\
$^{2}$European Southern Observatory, Karl-Schwarzschild-Stra{\ss}e 2, D-85748 Garching, Germany\\
$^{3}$Department of Astronomy, Box 351580, University of Washington, Seattle, WA 98195, USA\\
$^{4}$Smithsonian Astrophysical Observatory, 60 Garden St, Cambridge, MA 02138, USA\\
$^{5}$Institut de Recherche sur les Exoplan$\grave{e}$tes (iREx) and D$\acute{e}$partement de physique, Universit$\acute{e}$ de Montr$\acute{e}$al, Montr$\acute{e}$al, QC H3C 3J7, Canada\\
$^{6}$ Department of Physics and Astronomy, University College London, London WC1E 6BT, UK\\
$^{7}$Department of Physics, University of Warwick, Coventry CV4 7AL, UK\\
$^{8}$Department of Physics and Astronomy, University of Oklahoma, 440 W. Brooks St, Norman, OK 73019, USA\\
$^{9}$Department of Astronomy, Harvard University, 60 Garden St, Cambridge, MA 02138, USA\\
$^{10}$SETI Institute/NASA Ames Research Center, Moffet Field, CA 94035, USA
}

\date{Accepted XXX. Received YYY; in original form ZZZ}

\pubyear{2018}

\begin{document}
\label{firstpage}
\pagerange{\pageref{firstpage}--\pageref{lastpage}}
\maketitle

\begin{abstract}
Recent \textit{Kepler} photometry has revealed that about half of white dwarfs (WDs) have periodic, low-level ($\sim 10^{-4}-10^{-3}$), optical variations. \textit{Hubble Space Telescope} (\textit{HST}) ultraviolet spectroscopy has shown that up to about one half of WDs are actively accreting rocky planetary debris, as evidenced by the presence of photospheric metal absorption lines. We have obtained \textit{HST} ultraviolet spectra of seven WDs that have been monitored for periodic variations, to test the hypothesis that these two phenomena are causally connected, i.e. that the optical periodic modulation is caused by WD rotation coupled with an inhomogeneous surface distribution of accreted metals. We detect photospheric metals in four out of the seven WDs. However, we find no significant correspondence between the existence of optical periodic variability and the detection of photospheric ultraviolet absorption lines. Thus the null hypothesis stands, that the two phenomena are not directly related. Some other source of WD surface inhomogeneity, perhaps related to magnetic field strength, combined with the WD rotation, or alternatively effects due to close binary companions, may be behind the observed optical modulation.
We report the marginal detection of molecular hydrogen in WD\,J1949+4734, only the fourth known WD with detected H$_2$ lines. We also re-classify J1926+4219 as a carbon-rich He-sdO subdwarf.
\end{abstract}




\section{Introduction}
\label{sec:Intro}
Over 95~per cent of all stars end their lives as white dwarfs \citep[WDs;][]{Althaus_2010}. The observed properties of WDs thus hold the keys to numerous questions in diverse astronomical fields, from gravitational wave sources, through Type-Ia supernova progenitors, to the fate of planetary systems, to name a few. Studies have shown that many WDs accrete the debris of their former planetary systems \citep[e.g][]{Jura_2003, Zuckerman_2003, Zuckerman_2010}. This has emerged in parallel to the realization that planets and planetary systems are common around all stars \citep[e.g.][]{Winn_2015, Shvartzvald_2016}. WD debris accretion is sometimes evidenced in the discs detected around some WDs within the tidal disruption radius for typical asteroid densities. The discs appear as infrared-excess dust emission \citep[e.g.][]{Kilic_2006}, or as optical emission lines from gas \citep[e.g.][]{Manser_2016}. However, the most frequent manifestation of debris accretion is the photospheric ultraviolet (UV) metal absorption lines (typically Si, but sometimes C and other elements) found in $\approx 30-50$\,per cent of all WDs observed with the Cosmic Origins Spectrograph (COS) of the \textit{Hubble Space Telescope} (\textit{HST}) \citep{Gaensicke_2012, Koester_2014}.

The sinking timescales in the strong surface gravity of most WDs are relatively short. For example, in hydrogen-dominated-atmosphere (DA) WDs with effective temperatures between $12\,000$\,K and $25\,000$\,K, heavy elements would disappear from the WD atmosphere within weeks or less \citep{Koester_2006, Koester_2009a}. Thus, the presence of the metal lines indicates ongoing accretion of rocky material in at least $30$~per cent of all WDs \citep{Koester_2014}. The implied accretion rates are in the range $\sim 10^5 - 10^8$\,g\,s$^{-1}$ \citep{Koester_2014}. Furthermore, the \textit{HST} data have permitted detailed composition analyses, showing that the accreted material has a diverse make-up, but often a terrestrial-mantle-like composition \citep{Gaensicke_2012}, and sometimes even indications of large amounts of water \citep{Farihi_2013}. Debris-accreting WDs have thus opened a new window to study the properties of planetary material and its post-main-sequence fate \citep[e.g.][]{Xu_2017}.

The original \textit{Kepler} mission \citep{Borucki_2010} monitored more than $150\,000$ stars in a 115 deg$^2$ region for about four years, mainly in pursuit of stellar transits by Earth-like planets. Although the mission has focused primarily on main-sequence stars, a couple of dozen WDs were also observed, with unprecedented photometric precision. These WDs were included in the \textit{Kepler} sample as a part of the astroseismology survey \citep{Ostensen_2010}. Although the survey was aimed at pulsating compact objects (i.e. WDs and hot subdwarfs), WDs known to be outside of the instability strip, along with other WD candidates within the \textit{Kepler} field, were also included in the sample. However, none of the observed WDs turned out to be pulsating \citep{Ostensen_2011, Doyle_2017}. \citet{Maoz_2015} analysed the \textit{Kepler} time series for 14 WDs in the \textit{Kepler} database, and found periodic photometric modulations in seven of them. The variation periods were of order hours to 10 days, with amplitudes of order $10^{-3}$ to $10^{-4}$ , much lower than could have been detected with pre-\textit{Kepler} technology.\footnote{\textit{Kepler} data for an additional 13 WDs were analysed by \citet{Doyle_2017}. None of these 13 light curves show any periodic variability. From the total of 18 WDs with \textit{Kepler} data examined by \citet{Doyle_2017}, out of which five were included in the \citet{Maoz_2015} sample, only one (KIC\,11604781, or WD\,J1914+4936) displayed periodic variations, in agreement with \citet{Maoz_2015}. An additional WD (KIC\,8682822, or WD\,J1917+4452) classified as possibly variable at the low $60$\,ppm level by \citet{Maoz_2015}, was found to be non-variable by \citet{Doyle_2017}. After a re-analysis of its light curve we agree with \citet{Doyle_2017} and consider this WD non-variable, at the level of $\lesssim 60$\,ppm. Although, from the analysis of \citet{Doyle_2017} it might appear that the actual fraction of periodically variable WDs is lower than reported by \citet{Maoz_2015}, all of the additional 13 WDs analysed by \citet{Doyle_2017} have light curves of lower S/N, with upper limits of several hundred ppm on the variability -- and are hence less sensitive by an order of magnitude to the variations detected by \citet{Maoz_2015}. Many hundreds of additional WDs have been monitored in \textit{Kepler}'s \textit{K2} continuation campaigns, albeit with lower photometric precision and over few-month-long observing periods.}

In individual cases of these WDs, the observed modulation could possibly be explained by the effects of companions -- e.g. beaming \citep{Zucker_2007} due to reflex motion caused by a compact companion \citep[although such close double degenerates are rare;][]{Maoz_2017, Maoz_2018}, or reflection/re-radiation by a heated giant planet \citep[no planets around WDs have yet been discovered;][]{Fulton_2014} or a brown dwarf companion \citep[although no more than $\sim 2$~per cent of WDs have brown dwarf companions;][]{Girven_2011, Steele_2011}. However, the high occurrence rate of the modulation, and the typical periods, both suggest WD rotation \citep[e.g.][]{Hermes_2017} as the cause of the observed modulation. Furthermore, the fully radiative atmospheres of the warm WDs in the \textit{Kepler} sample argue also against variations caused by long-lived star spots, which are associated with photospheric convection that sets in only in cooler atmospheres \citep{Brinkworth_2005, Tremblay_2015}. Finally, the extremely high magnetic fields required for magnetic dichroism \citep{Angel_1981} make it unlikely that in more than one of the WDs the modulation is due to strong inhomogeneous surface magnetic fields. Instead, \citet{Maoz_2015} hypothesized that the optical-band low-level photometric modulation seen in one-half of WDs observed with \textit{Kepler} could be associated with the previously mentioned photospheric metal pollution, that is seen in one-half of WDs observed in the UV with the \textit{HST} \citep{Koester_2014}. Slightly inhomogeneous surface coverage of the accreted material (e.g. due to moderate magnetic fields) would lead to inhomogenous UV absorption. Optical fluorescence of the absorbed UV photons \citep{Pinto_2000}, combined with the WD rotation, could then potentially produce the observed levels and periods of optical modulation.

To test this hypothesis, we have obtained UV spectra of both variable and non-variable WDs. If our hypothesis is correct, there will be a one-to-one ``match'' between the WDs with/without modulations and the WDs with/without photospheric metal absorption lines.

\section{Observations}

\subsection{Sample selection}
\label{sec:Sample}
We have obtained UV spectra of seven WDs that have been monitored for low-level variability with a good signal-to-noise ratio (S/N): three WDs that show clear periodic modulations in the \textit{Kepler} data (or elsewhere), and four that do not, with upper limits on any periodic variability at amplitude levels below 100\,ppm (i.e. $10^{-4}$). To ensure sufficient S/N ($\gtrsim 10$ per resolution element), only WDs with a high expected UV signal were chosen. These were WDs with either a UV detection with \textit{GALEX} at $\approx 1550$\,\AA, or a visual magnitude extrapolated to the UV based on an effective temperature estimate such that desired S/N was expected. Additional selection criteria for all seven WDs were: effective temperatures $T_\textrm{eff} < 36\,000$\,K, to avoid WDs in which radiative levitation is dominant; and surface gravities $7.5 < \log g < 8.5$, to include only WDs of the kind whose UV metal pollution has been studied.

All four non-varying WDs were chosen from the \textit{Kepler} sample, since no other $\sim 100$\,ppm sensitivity level WD light curves exist. The variable WDs include two targets from \textit{Kepler}, and one additional target from the Pro-Am White Dwarf Monitoring (PAWM) survey\footnote{\href{http://brucegary.net/WDE/}{http://brucegary.net/WDE/}} \citep[WD\,2359$-$434, see Section~\ref{sec:WD2359} below;][]{Gary_2013}.

Our original sample included two additional objects: a non-varying WD (WD\,J1940+4240, see Section~\ref{sec:WD1940}), that turned out to be much cooler than previously estimated, resulting in no signal in the \textit{HST} far-UV (FUV) data; and a variable object (J1926+4219, see Section~\ref{sec:J1926}) that had been misclassified as a WD, and turned out to be a carbon-rich helium-dominated subdwarf. We exclude both of them from our analysis. Table~\ref{tab:Sample} lists the names, including the \textit{Kepler} Input Catalog (KIC) designation, and coordinates of all the nine objects from our original sample. In this paper we have adopted the naming convention of WD\,Jhh:mm$\pm$dd:mm based on the J2000 coordinates, unless a legacy B1950 WD\,hh:mm$\pm$dd:m name was already in use (i.e., WD\,1942+499 and WD\,2359$-$434).

The initial choice of nine WDs should have permitted, in principle, to significantly reject the null hypothesis that optical modulations and UV metal pollution are unrelated. Since the probability for any WD to display each effect alone is $\approx 50$~per cent \citep{Koester_2014, Maoz_2015}, the chance probability for $n$ matches would be $2^{-n}$. Even if in two of the nine WDs there could be a mismatch (due to a different cause for the modulation, such as rotation plus magnetic dichroism, or beaming caused by the presence of a close companion), this would still leave seven WDs with a match between the two effects. The chance probability for seven matches would be $2^{-7} < 1$~per cent, and thus seven or more matches would firmly reject the null hypothesis.

\begin{table}
\caption{Original sample: names and J2000 coordinates.}
\label{tab:Sample}
\begin{center}
\begin{tabular}{l l l l}
\hline
Name & KIC & RA & Dec\\
\hline
WD\,J1855+4207 & 6669882 & 18:55:46.033 & +42:07:04.43 \\
WD\,J1857+4909 & 11337598 & 18:57:47.150 & +49:09:38.60 \\
WD\,J1909+4717 & 10198116 & 19:09:59.347 & +47:17:10.00 \\
WD\,J1919+3958 & 4829241 & 19:19:27.685 & +39:58:39.71 \\
J1926+4219 & 6862653 & 19:26:46.011 & +42:19:35.32 \\
WD\,J1940+4240 & 7129927 & 19:40:59.367 & +42:40:31.30 \\
WD\,1942+499 & 11822535 & 19:43:43.686 & +50:04:37.80 \\
WD\,J1949+4734 & 10420021 & 19:49:14.579 & +47:34:45.98 \\
WD\,2359$-$434 & --- & 00:02:10.766 & -43:09:56.02 \\
\hline
\end{tabular}
\end{center}
\end{table}

\subsection{\textit{HST} Observations}
\label{sec:COSObs}

We obtained (\textit{HST} programme 14082, PI: Maoz) FUV spectra with COS \citep{Green_2012} for six WDs and the misidentified subdwarf, using the G130M grating with a central wavelength of $1291$\,\AA, which covers the wavelength range $1130-1435$\,\AA, excluding a gap between the two detector segments, at $1278-1288$\,\AA. The dispersion is $0.01$\,\AA\,pixel$^{-1}$. Each exposure was split into four sub-exposures, slightly dithered in the dispersion direction, in order to eliminate the COS MCD fixed-pattern noise. Four of the WDs were observed for one \textit{HST} orbit, while the fainter remaining three were observed for two orbits, in order to achieve a higher S/N. The observational setup is similar to that used by \citet{Koester_2014}, \citet{Farihi_2013}, \citet{Gaensicke_2012} in their search for UV metal absorption  lines in 85 WDs.

We retrieved archival data for two additional WDs: WD\,1942+499, for which a COS/FUV spectrum using a similar configuration was available from \textit{HST} programme 11526 (PI: Green); and WD\,2359$-$434, for which a Space Telescope Imaging Spectrograph (STIS) near-UV (NUV) spectrum was available from \textit{HST} programme 14076 (PI: G\"{a}nsicke). The STIS observations were performed using the G230L grating with a central wavelength of $2376$\,\AA, which covers the wavelength range $1600-3150$\,\AA, with a dispersion of about $1.5$\,\AA\,pixel$^{-1}$. All spectra underwent standard reduction by the appropriate \textit{HST} pipeline.

\subsection{Radial velocity measurements}
\label{sec:RVobs}
In order to rule out the presence of a massive WD or M-dwarf companion as a cause for the modulation in the variable WDs, and in general to examine the possibility of binarity in the sample, we have obtained ground-based spectroscopic observations, as described below. These observations are analysed and summarised in Section~\ref{sec:Companion}.

We used the Dual Imaging Spectrograph (DIS) mounted on the ARC 3.5-m telescope at Apache Point Observatory (APO) in New Mexico on the nights of 2017 April 30 and 2017 June 17. We obtained a total of 27 epochs of the seven northern objects in our sample, with an exposure time varying between 300 and 1200\,s. We used the 1200\,lines\,mm$^{-1}$ gratings centred around $4380$\,\AA\ and $6500$\,\AA, and a 0.9\,arcsec slit, covering the range 3750 to 5010\,\AA\ with 0.62\,\AA\,pix$^{-1}$ dispersion in the blue channel, and 5920 to 7100\,\AA\ with 0.56\,\AA\,pix$^{-1}$ dispersion in the red channel. To improve our wavelength calibration, a HeNeAr comparison lamp spectrum was obtained after each observation. The weather was clear on both nights, yielding a S/N of $\approx 3-5$ for the first night and $\approx 2-15$ for the second night. Standard \textsc{iraf} packages were used to perform bias correction, flat-fielding, spectrum extraction, and wavelength calibration. Flux calibration was completed using standard \textsc{iraf} packages with the ESO spectrophotometric standard star BD\,+33d2642.

Eight additional epochs of five of the northern objects were obtained using the 6.5-m MMT telescope Blue Channel Spectrograph on Mt. Hopkins, Arizona, on the nights of 2017 June 24-26. We used the 832\,lines\,mm$^{-1}$ grating in the second order and a 1\,arcsec slit, covering the range 3570 to 4510\,\AA\ with 0.36\,\AA\,pix$^{-1}$ dispersion and a 1\,\AA\ spectral resolution. The exposure time varied between 75 and 720\,s. Wavelength calibration was performed with arc lamp spectra obtained immediately after each exposure. Because the targets are relatively bright, the observations could be made during poor seeing and/or some cloud cover. The spectra have a S/N per resolution element between $20$ to $50$ in the continuum.

\section{Results and analysis}
\label{sec:Results}
Figs~\ref{fig:WD1909}--\ref{fig:WD2359} show the \textit{HST} pipeline-reduced spectra for each of the seven WDs in the observed sample. We have visually searched each spectrum for absorption lines. To determine the nature of the lines (photospheric or interstellar), we have fitted each spectrum with a synthetic WD model generated with the spectral synthesis program \textsc{Synspec} \citep[version 50;][]{Hubeny_2011}, based on model atmospheres created by the \textsc{Tlusty} program \citep[version 205;][]{Hubeny_1988, Hubeny_1995, Hubeny_2017a}. \textsc{Tlusty} produces one-dimensional, horizontally homogeneous, plane-parallel model atmospheres in hydrostatic equilibrium. Following the moderately cool WD example of \citet{Hubeny_2017b}, we have assumed ML2-type convection with a mixing length of 0.6 of a pressure scale height. The models were computed in local thermodynamic equilibrium, using the Tremblay tables \citep{Tremblay_2009} for hydrogen line broadening.

For each WD, the basic model atmosphere was computed with \textsc{Tlusty} using the published effective temperature and surface gravity estimates, except in the case of WD\,J1949+4734 and J1926+4219, where the published values were inconsistent with the observations (see below). We then used \textsc{Synspec} to produce various synthetic spectra with different Si and C abundances, in steps of 1\,dex, for each element separately. Each model spectrum was convolved with the COS line spread function\footnote{\href{http://www.stsci.edu/hst/cos/performance/spectral\_resolution}{http://www.stsci.edu/hst/cos/performance/spectral\_resolution}}. The model was then scaled and Doppler shifted to match the observed spectrum, using $\chi^2$ minimization. Following a visual inspection of the observed spectrum and the various model spectra, we repeated the procedure using smaller steps of 0.2\,dex, around the best-fitted abundance value. Finally, a combined model spectrum consisting of both Si and C with the matched abundances was created, convolved, shifted, and scaled to fit the observed spectrum, or to find the upper limits on metal abundances in the cases with no detection of either or both elements. We note that the various expected metal absorption line strengths depend strongly on both temperature and surface gravity, and therefore the levels of the abundance limits in the cases of non-detection vary among the WDs. The results, including the derived abundances, are summarised in Table~\ref{tab:Results}. Figs~\ref{fig:WD1909}--\ref{fig:WD2359} also show the model fits in selected regions. The absorption lines absent from the models are of interstellar origin (interstellar lines of \ion{C}{ii}, \ion{N}{i}, \ion{O}{i}, \ion{Si}{ii}, \ion{S}{ii}, and \ion{Fe}{ii} were detected), unless mentioned otherwise. The results for each WD are discussed in detail below.

In their \textit{HST} survey of warm ($17\,000-27\,000$\,K) DA-type WDs,  \citet{Koester_2014} detected photospheric metal absorption lines in 48 out of 85 WDs. All of these 48 WDs had Si lines in their atmosphere at abundances of $-8.5 \lesssim$ \abd{Si} $\lesssim -5$ (all abundances are by number). In 18 of them C lines were also detected ($-8.5 \lesssim$ \abd{C} $\lesssim -5.5$), and seven more had other metals as well (Mg, Al, P, S, Ca, Cr, Fe, or Ni). For reference, solar abundances are \abd{Si} $=-4.49 \pm 0.03$ and \abd{C} $=-3.57 \pm 0.05$, see \citealt{Asplund_2009}. Since Si has always been detected when any metals are present in the sample of \citet{Koester_2014}, and because our upper limits on C in some WDs are comparable to the detected levels in other WDs, we will generally use the Si detection to determine if a WD has atmospheric metals or not. However, in WD\,J1949+4734 we have a clear detection of C at a relatively high abundance without Si (see Section~\ref{sec:WD1949} below), and we hence count it as a case of metal line detection.

\begin{table*}
\caption{WD properties --- effective temperature ($T_{\textrm{eff}}$), surface gravity ($\log g$), optical variability period ($P$), amplitude ($A$); and Si and C abundances relative to hydrogen, as determined from the model fits to the \textit{HST} data. The `Z?' column indicates detection of photospheric metal lines. The `Hyp?' column indicates whether the WD matches our experiment's hypothesis, that optical variability and debris accretion are related. See Section~\ref{sec:Results} for further details.}
\label{tab:Results}
\begin{center}
\begin{tabular}{l l c c c c c c c c c c}
\hline
Name & Spectral & $T_{\textrm{eff}}$ & $\log g$ & $P$ & $A$ & \abd{Si} & \abd{C} & Z? & Hyp? & Notes & Ref. \\
 & type & [kK] & & [h] & [ppm] & & & & & & \\
\hline
\multicolumn{11}{l}{Non-variable WDs:}\\
WD\,J1909+4717 & DA & 13.5 & 8.0 & -- & $<70$ & $-7.8\pm0.2$ & $<-8.0$ & + & -- & 1, 2 & a, e \\
WD\,J1919+3958 & DA & 19.5 & 8.0 & -- & $<50$ & $<-9.0$ & $<-8.5$ & -- & + & & a, e\\
WD\,1942+499 & DA & 36 & 7.9 & -- & $<20$ & $-6.3\pm0.3$ & $<-8.5$ & + & -- & 3 & b, e \\
WD\,J1949+4734 & DA & 12.8 & 7.8 & -- & $<50$ & $<-8.0$ & $-6.7\pm0.3$ & + & -- & 2, 4 & c, e \\
\hline
\multicolumn{11}{l}{Periodically variable WDs:}\\
WD\,J1855+4207 & DA & 30.5 & 7.4 & $8.81\pm0.22$ & 800 & $-7.4\pm0.2$ & $-7.8\pm0.2$ & + & + & 3 & b, e \\
WD\,J1857+4909 & DA & 22.8 & 8.6 & $2.23872\pm0.00072$ & 300 & $<-8.5$ & $<-8.5$ & -- & -- & & b, e\\
WD\,2359$-$434 & DA & 8.6 & 8.3 & $2.694926\pm0.000073$ & 5000 & $<-9.5$ & $<-8.0$ & -- & ? & 5, 6 & d, f \\
\hline
\end{tabular}
\end{center}

\begin{flushleft}
Notes:
(1) Tentative Si detection.
(2) Low S/N.
(3) Possible detection of circumstellar material.
(4) H$_2$ detected.
(5) NUV spectrum only.
(6) \abd{Mg} $< -10$.

References:
$T_{\textrm{eff}}$ and $\log g$ based on (a) \citet{Doyle_2017}, (b) \citet{Ostensen_2011}, (c) this work (see text), (d) \citet{Giammichele_2012}.
$P$ and $A$ based on (e) \citet{Maoz_2015}, or (f) \citet{Gary_2013}.
\end{flushleft}

\end{table*}

\subsection{Non-variable WDs}

\begin{figure}
\centering
\includegraphics[width=\columnwidth]{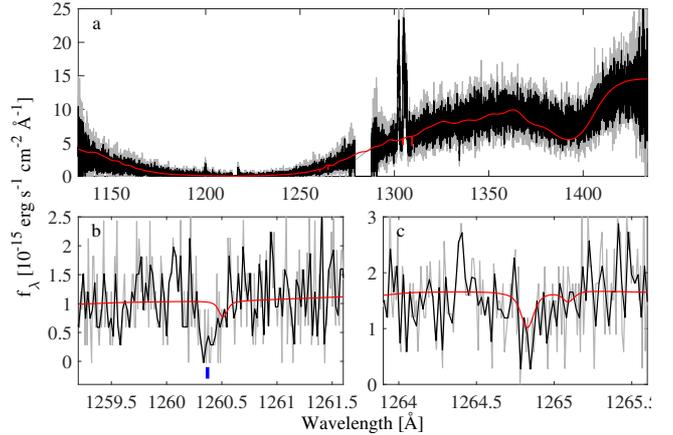}
\caption{(a) \textit{HST}/COS spectrum of WD\,J1909+4717 (unbinned, gray; binned every two data points, black) and model fit (red). Unmodelled absorption lines are of interstellar origin. Airglow of \ion{O}{i} is visible around $1302-1306$\,\AA. (b) Interstellar \ion{Si}{ii} 1260.4\,\AA\ line (blue tick) with a null or weak detection of the photospheric component of this line in the red wing of the interstellar line (red model). (c) Tentative detection of a photospheric \ion{Si}{ii} 1264.7\,\AA\ line.}
\label{fig:WD1909}
\end{figure}

\paragraph*{WD\,J1909+4717}
\label{sec:WD1909}
Although this WD was observed for two \textit{HST} orbits, the obtained COS/FUV spectrum is still rather noisy, with S/N $\approx 2$ (or $\approx 3$ after binning) at the relevant wavelength range. Nevertheless, we report a tentative detection of a photospheric \ion{Si}{ii} 1264.7\,\AA\ line, at an abundance of \abd{Si} $\approx -7.8$ (see Fig.~\ref{fig:WD1909}). Since under photospheric conditions the $1265$\,\AA\ line is always stronger than the $1260$\,\AA\ line \citep{Koester_2014}, its presence, if real, indicates a photospheric origin. The photospheric origin is also independently supported by the weighted mean RV measured from the Balmer lines in the optical spectra (see Section~\ref{sec:Companion}), that is consistent with the RV of the candidate photospheric absorption line ($\approx 20$\,km\,s$^{-1}$). An interstellar $1260$\,\AA\ \ion{Si}{ii} line, shifted by $\approx 40$\,km\,s$^{-1}$ blue-ward compared to the photospheric line (see Fig.~\ref{fig:WD1909}b), is also visible. If this is a true detection of a photospheric metal line, it contradicts the prediction from our hypothesis, of no photospheric metals in non-variable WDs.

The UV spectrum of this relatively cool DA shows the wide $1400$\,\AA\ Ly$\alpha$ satellite line from collisions of H--H$^+$, common in DA WDs cooler than $20\,000$\,K \citep{Greenstein_1979, Wegner_1982, Koester_1985, Nelan_1985}.

\begin{figure}
\centering
\includegraphics[width=\columnwidth]{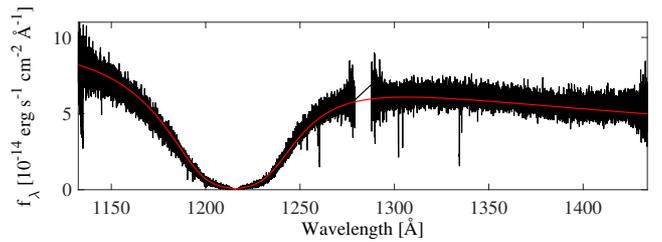}
\caption{\textit{HST}/COS spectrum of WD\,J1919+3958 (black) and model fit (red). Unmodelled absorption lines are of interstellar origin. No photospheric metal lines are detected within our sensitivity limits.}
\label{fig:WD1919}
\end{figure}

\paragraph*{WD\,J1919+3958}
No photospheric metal lines were detected in this WD (see Fig.~\ref{fig:WD1919}), in accord with expectations from our hypothesis.

\begin{figure}
\centering
\includegraphics[width=\columnwidth]{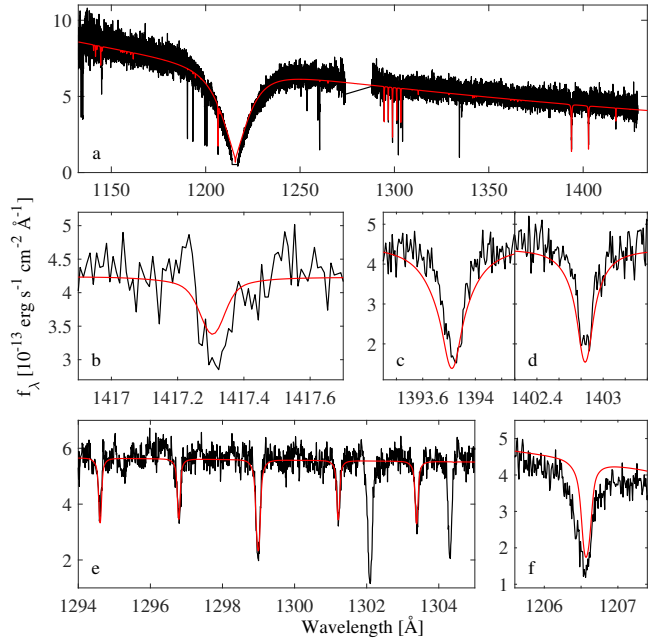}
\caption{(a) \textit{HST}/COS spectrum of WD\,1942+499 (black) and model fit (red). Unmodelled absorption lines are of interstellar origin. (b) \ion{Si}{iii}. (c)+(d) \ion{Si}{iv}. (e) \ion{Si}{iii}. The two strong lines absent from the model are interstellar \ion{O}{i} $1302.2$\,\AA\ and \ion{Si}{ii} $1304.4$\,\AA. (f) \ion{Si}{iv} $1206.5$\,\AA\ absorption line. Note the slight discrepancy between the model fit and the observed line, which may have an additional blue-shifted component, perhaps of circumstellar origin.}
\label{fig:WD1942}
\end{figure}

\paragraph*{WD\,1942+499}
\label{sec:WD1942}
Since for this WD we use archival data obtained in 2010, in our modelling we have used Lifetime Position 1 of the COS line spread function, as opposed to Lifetime Position 3 used for the other COS observations. Si absorption lines from excited states that cannot have originated in the interstellar medium (\ion{Si}{iii} and \ion{Si}{iv}) are detected in the COS/FUV spectrum (see Fig.~\ref{fig:WD1942}), with our modelling indicating an abundance of \abd{Si} $\approx -6.3$. The RV estimated from the Si lines is consistent with the Balmer-line RV measured from the optical spectra, supporting the case for the photospheric origin of these  absorption lines. Based on the lack of \ion{C}{ii} or \ion{C}{iii} absorption lines in the spectrum, we were able to constrain the C abundance to \abd{C} $<-8.5$. Nevertheless, in a previous study by \citet{Lallement_2011} of the same COS data along with an additional G160M-grating spectrum (covering the wavelength range $1400-1750$\,\AA), \ion{C}{iv} $1548$ and $1551$\,\AA\ absorption lines were detected, at an abundance of \abd{C} $\sim -7.6$. These lines may have a slight blue-shift relative to that of photospheric lines, but the shift is also consistent with zero within the measurement error \citep{Barstow_2010, Lallement_2011}. These authors also note a discrepancy in the Si abundance, as measured from the \ion{Si}{iii} (\abd{Si} $\sim -6.0$) and \ion{Si}{iv} (\abd{Si} $\sim -6.7$) lines. \citet{Lallement_2011} have suggested that the high-ionization absorption lines (\ion{Si}{iv}, \ion{C}{iv}, and \ion{N}{v}, which is marginally detected) originate from circumstellar material surrounding the WD, and not from the photosphere. If these lines are circumstellar, then it is puzzling that they are not blue-shifted by $\sim 25-30$\,km\,s$^{-1}$ relative to the photospheric Balmer lines, which undergo gravitational redshift. Indeed, in \citet{Debes_2012} a claimed circumstellar \ion{Ca}{ii} line component in WD\,1124$-$293 is blue-shifted by $30$\,km\,s$^{-1}$ relative to the main photospheric component, as one would expect. In the present case, this would suggest that either the circumstellar material is in radial infall at roughly such a velocity, or that the material is only slightly above the photosphere, such that the lines undergo a similar gravitational redshift (a similar phenomenon is presented and discussed below, Section~\ref{sec:WD1855}, for WD\,J1855+4207). Even if the high-ionization absorption lines do have a circumstellar origin, the \ion{Si}{iii} lines (and possibly part of the \ion{Si}{iv} material, see Fig.~\ref{fig:WD1942}(f)) are consistent with a photospheric model fit. Thus, although some of the material might be circumstellar, metals are present also in the WD photosphere.

Metals from a past accretion event can be supported in the atmospheres of hot DAs such as this one ($T_\textrm{eff}\approx 36\,000$\,K) by radiative levitation, leaving enough time for the metal distribution to become homogeneous, through transverse diffusion, over the WD surface. If so, the Si detection in a non-variable WD does not necessarily contradict the expectation from our hypothesis. However, the detected Si abundance is higher than expected from radiative levitation \citep{Chayer_1995}, and therefore cannot be explained solely by a past accretion event. We thus count this case as a mismatch to our hypothesis.

\begin{figure}
\centering
\includegraphics[width=\columnwidth]{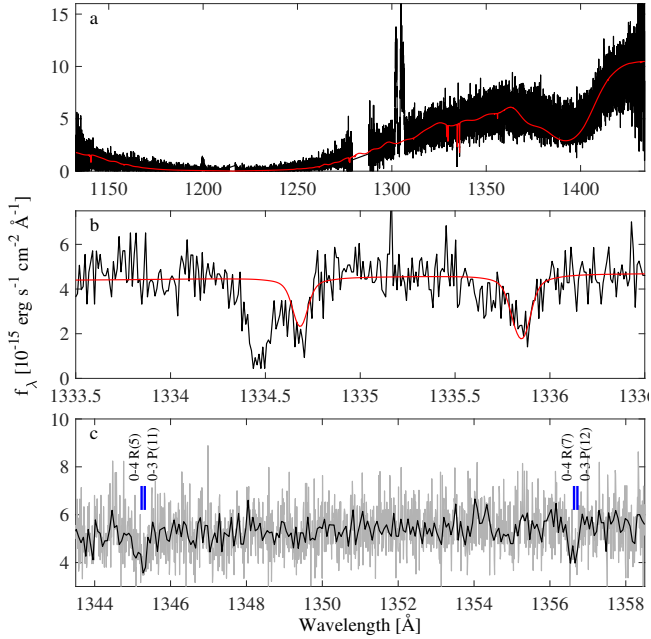}
\caption{(a) \textit{HST}/COS spectrum of WD\,J1949+4734 (black) and model fit (red). Unmodelled absorption lines are of interstellar origin. Airglow of \ion{O}{i} is visible around $1302-1306$\,\AA. (b) Interstellar and photospheric \ion{C}{ii} lines. (c) Molecular hydrogen absorption lines (unbinned spectrum, gray; binned every six data points, black). The blue ticks mark the theoretical wavelengths of the H$_2$ 0--4\,R(5), 0--3\,P(11), 0--4\,R(7), and 0--3\,P(12) transitions \citep{Abgrall_1993, Xu_2013}, redshifted by $\approx 30$\,km\,s$^{-1}$ to the WD rest frame.}
\label{fig:WD1949}
\end{figure}

\paragraph*{WD\,J1949+4734}
\label{sec:WD1949}
Two \textit{HST} orbits were spent on this WD. Despite the rather low S/N of the obtained COS/FUV spectrum, a pair of $1334$/$35$ \ion{C}{ii} absorption lines at an abundance of \abd{C} $\approx -6.7$ are clearly detected (see Fig.~\ref{fig:WD1949}). The photospheric origin of these lines is established in this case by the detection of an interstellar $1334$\,\AA\ \ion{C}{ii} line, shifted blue-ward by $\approx 50$\,km\,s$^{-1}$ compared to the photospheric line, and the RV consistency with the optical Balmer lines.

Although \citet{Ostensen_2011} estimated $T_\textrm{eff} \approx 16\,200$\,K and $\log g \approx 7.8$ based on the optical spectrum of this WD, our fit to the UV spectrum suggests a significantly lower effective temperature of $\approx 12\,750$\,K. To find the best-fitting model, we varied the $T_\textrm{eff}$ around the value estimated from the spectral energy distribution (SED), until a sufficient fit was achieved. Similarly to WD\,J1909+4717 (see Section~\ref{sec:WD1909}), this WD also shows the wide $1400$\,\AA\ Ly$\alpha$ satellite line from collisions of H--H$^+$. In addition, several molecular hydrogen lines are marginally detected in the WD rest frame (see Fig.~\ref{fig:WD1949}c), making it only the fourth known WD with molecular hydrogen (see \citealt{Xu_2013} for the first three). This detection of H$_2$ is consistent with our lower effective temperature estimate for this WD (see \citealt{Zuckerman_2013} for the similar case of GD\,31).

Unless the presence of C in the atmosphere is the result of carbon dredge-up due to convection, this WD also constitutes a mismatch, counter to expectations from our hypothesis.

\subsection{WDs with periodic modulations}
\begin{figure}
\centering
\includegraphics[width=\columnwidth]{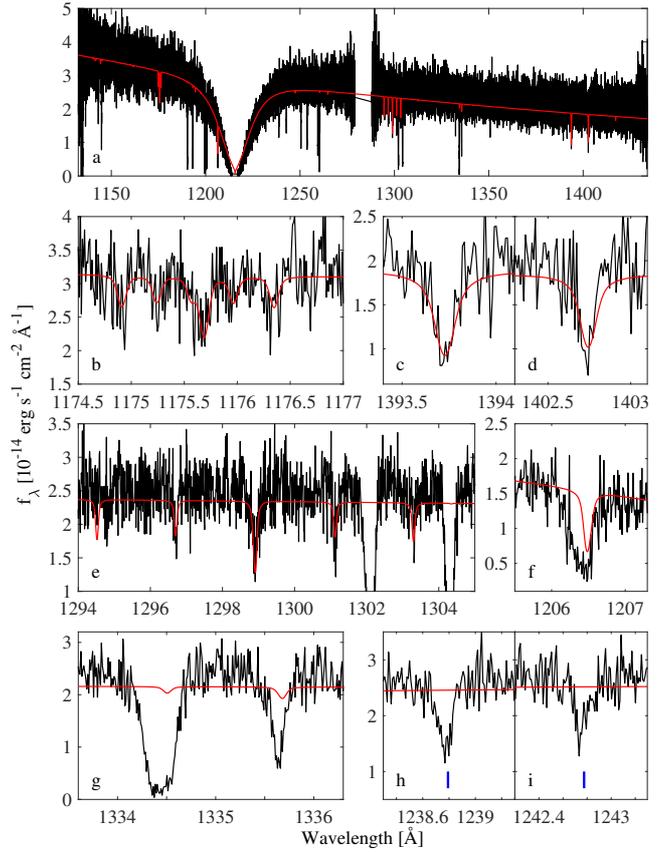}
\caption{(a) \textit{HST}/COS spectrum of WD\,J1855+4207 (black) and model fit (red). Unmodelled absorption lines are of interstellar origin. (b) \ion{C}{iii}. (c)+(d) \ion{Si}{iv}. (e) \ion{Si}{iii}. The strong unmodelled lines are interstellar \ion{O}{i} $1302.2$\,\AA\ and \ion{Si}{ii} $1304.4$\,\AA. (f) \ion{Si}{iv} $1206.5$\,\AA\ absorption line. Note the discrepancy between the model fit and the observed line, which has an additional blue-shifted component, possibly circumstellar. (g) Interstellar \ion{C}{ii} lines, with some small contribution from the photosphere. (h)+(i) \ion{N}{v} absorption lines, possibly from circumstellar material (see Section~\ref{sec:WD1855}). The blue ticks mark the WD rest frame.}
\label{fig:WD1855}
\end{figure}

\paragraph*{WD\,J1855+4207}
\label{sec:WD1855}
The COS/FUV spectrum of this WD (Fig.~\ref{fig:WD1855}) shows photospheric absorption lines from \ion{C}{ii}, \ion{C}{iii}, \ion{Si}{iii}, and \ion{Si}{iv}, in agreement with expectations from our hypothesis. The RV of the photospheric lines is broadly consistent with that measured from the low S/N optical spectra of this object. However, as in the case of WD\,1942+499 (see Section~\ref{sec:WD1942}), there is a discrepancy in the Si abundance measured from \ion{Si}{iii} and \ion{Si}{iv} (see Fig.~\ref{fig:WD1855}(f), for example). High-ionization \ion{N}{v} $1239$\,\AA\ and $1243$\,\AA\ absorption lines are also detected, close to the photospheric velocity (Fig.~\ref{fig:WD1855}(h+i)). These \ion{N}{v} lines cannot have originated from the $\sim 30\,500$\,K photosphere, and require a temperature $\gtrsim 80\,000$\,K \citep{Sion_1998, Long_1999}. Thus, WD\,J1855+4207 seems to have both photospheric and possibly infalling circumstellar metals, much like WD\,1942+499.

The cases of WD\,J1855+4207 and WD\,1942+499, discussed above, both of them hot $T_\textrm{eff} > 30\,000$\,K objects with indications of highly ionized circumstellar metals, in addition to the detected photospheric metals, are similar to the case of GD\,394 (WD\,2111+489), a $T_\textrm{eff} \sim 35\,000$\,K DA WD showing high-ionization \ion{C}{iv}, \ion{N}{v}, and \ion{P}{v} lines, all approximately at the photospheric, gravitationally redshifted, velocity \citep[][in preparation]{Chayer_2000, Wilson_2018}. Closely related may be also the cases of SDSS\,1228+1040 and WD\,0843+516 \citep{Koester_2014}, that show high-excitation \ion{Si}{iv} absorption (but no \ion{N}{v}), that is again slightly blue-shifted relative to the photospheric lines. This may be an emerging class of hot, polluted, WDs with signs of a circumstellar envelope that is either infalling, or perhaps levitated just above the photosphere.  

\begin{figure}
\centering
\includegraphics[width=\columnwidth]{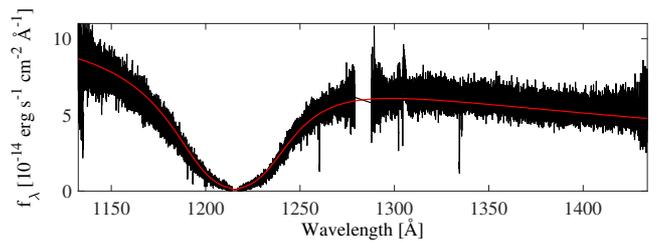}
\caption{\textit{HST}/COS spectrum of WD\,J1857+4909 (black) and model fit (red). Unmodelled absorption lines are of interstellar origin. Airglow of \ion{O}{i} is visible around $1302-1306$\,\AA. No photospheric metal lines are detected within our sensitivity limits.}
\label{fig:WD1857}
\end{figure}

\paragraph*{WD\,J1857+4909}
No photospheric metal absorption lines are detected (see Fig.~\ref{fig:WD1857}), making this case a mismatch to expectations from our hypothesis.

\begin{figure}
\centering
\includegraphics[width=\columnwidth]{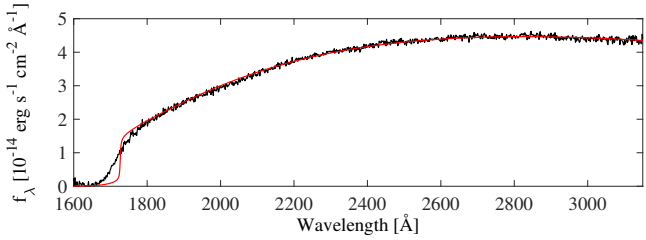}
\caption{\textit{HST}/STIS spectrum of WD\,2359$-$434 (black) and model fit (red). No photospheric metal lines are detected within our sensitivity limits.}
\label{fig:WD2359}
\end{figure}

\paragraph*{WD\,2359$-$434}
\label{sec:WD2359}
This WD was not part of the \textit{Kepler} sample, but was discovered to have periodic modulations by the PAWM survey\footnote{\href{http://brucegary.net/WDE/WD2359-434/WD2359-434.htm}{http://brucegary.net/WDE/WD2359-434/WD2359-434.htm}} \citep{Gary_2013}. It was observed using STIS and we have therefore used the STIS line spread function\footnote{\href{http://www.stsci.edu/hst/stis/performance/spectral\_resolution}{http://www.stsci.edu/hst/stis/performance/spectral\_resolution}} to broaden the model spectrum. We have also checked the NUV spectrum of this WD for the presence of Mg, since strong \ion{Mg}{i} absorption lines are expected around $2800$ and $2850$\,\AA. Nevertheless, the spectrum shows no absorption lines of any kind (see Fig.~\ref{fig:WD2359} and Table~\ref{tab:Results}), suggesting a mismatch to our hypothesis. However, since this WD is cool enough to be convective, the observed modulation could be the result of normal star spots (i.e. photospheric regions in which weak magnetic fields restrain the convective circulation, resulting in lower effective temperatures), as already proposed by \citet{Gary_2013}. We note that \citet{Gary_2013}'s alternative explanation for the modulation, of reflection from an orbiting giant planet, is unlikely because the planet would be well within the tidal disruption radius for a gas planet around such a WD. Reflection from a cool ($\lesssim 1000$\,K) brown dwarf companion is a possibility (see Section~\ref{sec:Companion}).

The possibility of a moderate magnetic field in this WD was already raised by \citet{Koester_1998}, based on the flat H$\alpha$ core shape, and confirmed by \citet{AznarCuadrado_2004} and later by \citet{Kawka_2007} using spectropolarimetry, yielding a longitudinal field strength of $\approx 3$\,kG. A mean field strength of $\approx 110$\,kG was measured by \citet{Koester_2009b}, based on high-resolution spectroscopy. A recent study by \citet{Landstreet_2017} monitored the variations in the magnetic field of this WD, based on the equivalent width of the Zeeman-split H$\alpha$ line core, and found a period similar to the one of the photometric modulations detected by \citet{Gary_2013}. \citet{Landstreet_2017} suggested that the variations observed in the Zeeman splitting are the result of the WD rotation combined with a non-axissymmetric multipolar magnetic field.

The UV spectrum of this DA shows the wide $1600$\,\AA\ Ly$\alpha$ satellite line from collisions of H--H, common in DA WDs cooler than $13\,500$\,K, as already observed by \citet{Koester_1985}. As can be seen in Fig.~\ref{fig:WD2359}, our \textsc{Synspec} model does not precisely reproduce the shape of the $1600$\,\AA\ feature. This could be the result of suppressed convection as the result of the magnetic field \citep{Tremblay_2015, GentileFusillo_2017}, leading to a partially-radiative atmosphere, that is inconsistent with the convective model that we use. However, if the magnetic field in this WD fully suppresses the convection on the WD surface, then it is hard to see how the star spots, presumably behind the optical modulations, could form. It is thus difficult to explain both the modulations and the spectral mismatch by means of the magnetic field.

\subsection{Excluded objects}
\paragraph*{WD\,J1940+4240}
\label{sec:WD1940}
The \textit{Kepler} light curve of this WD shows no modulations down to $<80$\,ppm \citep{Maoz_2015}. Based on a poor atmospheric model fit, yielding $T_\textrm{eff} \approx 23\,000$\,K, \citet{Ostensen_2011} conjectured that this may be a DA+DA double WD. Unfortunately, our 2-orbit-long COS observation resulted in almost no signal, as would have been expected from the recent, much cooler (and hence UV faint) $T_\textrm{eff} \approx 9\,500$\,K estimate of \citet{Doyle_2017}. As already noted, we exclude this WD from our analysis.

\begin{figure}
\centering
\includegraphics[width=\columnwidth]{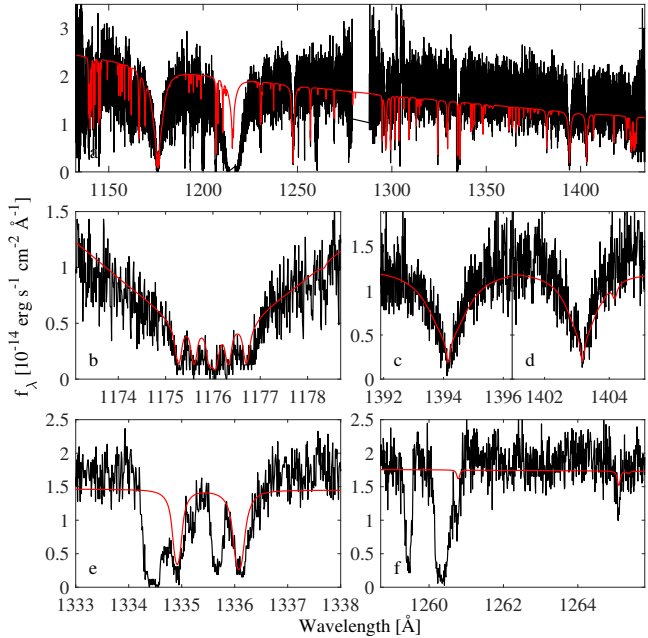}
\caption{(a) \textit{HST}/COS spectrum of the hot subdwarf J1926+4219 (black) and model fit (red). Unmodelled absorption lines are of interstellar origin. (b) \ion{C}{iii}. (c)+(d) \ion{Si}{iv}. (e) Photospheric and unmodelled interstellar \ion{C}{ii} lines. (f) Photospheric \ion{Si}{ii} lines. The strong unmodelled lines are interstellar \ion{S}{ii} $1259.5$\,\AA\ and \ion{Si}{ii} $1260.4$\,\AA\ lines.}
\label{fig:WD1927}
\end{figure}

\paragraph*{J1926+4219}
\label{sec:J1926}
This object was classified as a helium-dominated-atmosphere (DB) WD by \citet{Ostensen_2011}. They estimated, based on line fits to the optical spectrum, $T_\textrm{eff} \approx 16\,000$\,K, but reported a poor fit. The \textit{Kepler} light curve shows $14.16\pm0.48$\,h, $500$\,ppm, periodic variations \citep{Maoz_2015}. From its SED based on available photometry we estimated, prior to our observations, an effective temperature of $\approx 34\,000$\,K, and a high S/N in the UV that we indeed obtained. Photospheric \ion{C}{ii}, \ion{C}{iii}, \ion{Si}{iii}, and \ion{Si}{iv} absorption lines are detected in the FUV spectrum (see Fig.~\ref{fig:WD1927}). However, during the modelling process we realized that this is in fact a helium-rich hot subdwarf (He-sdO). Using the \textsc{Tlusty}-calculated NLTE (non-local thermodynamic equilibrium) grid of hot subdwarfs of \citet{Fontaine_2014} to fit the co-added optical MMT spectrum we have acquired (Section~\ref{sec:RVobs}), we estimate $T_\textrm{eff}=40\,300 \pm 210$\,K, $\log g = 6.40\pm0.11$, and \abd{He}$=3.17\pm0.79$. We then use these parameters to create an NLTE \textsc{Tlusty} model with C and Si, to fit the \textit{HST} UV spectrum, and we find abundances of $\log \left(\textrm{C}/\textrm{He}\right) = -2.5\pm0.3$ and $\log \left(\textrm{Si}/\textrm{He}\right)=-4.6\pm0.2$. The high C abundance is consistent with the reported abundances of carbon-enhanced He-sdOs \citep{Heber_2016}, although most of these reports have been for subdwarfs with a relatively higher effective temperature ($T_\textrm{eff}>43\,300$\,K), while only two out of the 16 reported carbon-enhanced He-sdOs have $T_\textrm{eff}\approx40\,000$\,K (like J1926+4219). Subdwarfs are evolved stars whose envelopes have been stripped, and therefore must have, or must have had in the past, a close companion. We discuss in Section~\ref{sec:Companion} the constraints on the presence of a binary companion to this subdwarf.

\section{A search for binary companions}
\label{sec:Companion}

In order to test for a companion-induced (rather than rotation related) cause for the optical modulation of some of the WDs, we searched for the presence of companions using two approaches. First, we searched the SED for evidence of a cool stellar or sub-stellar component. We examined on VizieR \citep{Ochsenbein_2000} the available optical photometry of each of the WDs, and compared it to a Planck spectrum with a temperature of the atmospheric fit listed in Table~\ref{tab:Results}, scaled to fit the optical-band photometry of the WD. We then searched for evidence of any near-infrared (NIR) excess at $3.6-8\,\mu$m as indicated by \textit{Spitzer Space Telescope} IR Array Camera (IRAC) photometry available from the \textit{Spitzer} Heritage Archive. For WD\,1942+499 and WD\,2359$-$434 we used cryogenic $4.5\,\mu$m and $8.0\,\mu$m images from the ``survey for planets and exozodiacal dust around WDs'' \citep[\textit{Spitzer} programme 2313;][]{Kuchner_2004,Mullally_2007}. We used available $3.6\,\mu$m and $4.5\,\mu$m images from the warm ``Small \textit{Spitzer} \textit{Kepler} Survey'' \citep[Small SpiKeS; \textit{Spitzer} programme 90100;][]{Werner_2012} for J1926+4219, and from SpiKeS \citep[\textit{Spitzer} programme 10067;][]{Werner_2013} for the remaining WDs. The photometry was performed using \textsc{mopex} \citep{Makovoz_2005} Point Response Function (PRF) fitting, corrected for the wavelength-dependent sensor sensitivity using the Rayleigh-Jeans tail correction factors provided by \citet{Reach_2005}. To quantify any detected excesses and estimate the detection limits for excesses, to each model Planck spectrum (representing the WD), we added an M-dwarf model \citep{Allard_2016} of varying temperature, scaled according to the ratio of the surface areas of the WD and of an M-dwarf companion. The WD radii were estimated using the effective temperature and surface gravity listed in Table~\ref{tab:Results}, and the theoretical WD cooling tracks of \citet{Fontaine_2001}\footnote{\url{http://www.astro.umontreal.ca/~bergeron/CoolingModels/}}. We find no evidence for a companion warmer than $\approx 1000$\,K for all the WDs in our sample, except for the case of WD\,J1855+4207 in which the WD is too faint to be detected in the IR, and hence we can only exclude companions warmer than $\approx 2500$\,K. Similarly, we have no \textit{Spitzer} detection for the hot subdwarf J1926+4219, and so can exclude only companions warmer than $\approx 4000$\,K.

Second, we have tested for massive stellar companions by searching for RV variations in the ground-based spectroscopic data, described in Section~\ref{sec:RVobs}. We estimated the RV of each epoch by cross-correlating the observed spectrum with a synthetic model spectrum generated by \textsc{Tlusty} using the atmospheric parameters that are listed in Table~\ref{tab:Results}. The measured velocities were corrected to the barycentre rest frame using the \textsc{earth\_vel\_ron\_vondrak} \textsc{Matlab} function of \citet{Ofek_2014}. The RV uncertainty, estimated from the distribution of the RV differences between epochs of the same objects (scaled down by a factor $\sqrt{2}$ to account for the two measurement errors that enter a difference), was typically around $\approx 25$\,km\,s$^{-1}$, but could also be as large as $\approx 80$\,km\,s$^{-1}$ in epochs with a low S/N. No significant RV variations are detected, within the limits of our sensitivity. We examine the implications of these RV variation limits for a putative binary companion for each of the periodically variable WDs in the sample.

For WD\,J1855+4207, with a primary mass of about $0.45$\,\msun\ \citep[based on the cooling tracks of][]{Fontaine_2001} and an orbital period of $8.8$\,h, the upper limit on any companion mass, $M_2$, is $M_2 \sin i \lesssim 0.20$\,\msun, where $i$ is the orbital inclination angle. The relatively low mass of this WD places it at the lower limit of single-star evolution \citep[single stars that evolve into $\lesssim 0.45$\,\msun\ WDs have not left the main sequence yet; e.g.][]{Brown_2016}. Beaming modulation of the WD flux due to reflex motion caused by a massive cool companion (e.g. a cool WD) requires $M_2 \sin i>1$\,\msun\ \citep{Maoz_2015} and can be ruled out. Modulation due to reflection of the WD light by a giant planet is also unlikely since the planet would be marginally within the WD's tidal radius for its density. The relatively high upper limit on the possible companion effective temperature ($\approx 2500$\,K), does not allow us to rule out reflection or re-radiation from a brown dwarf companion. A remaining possible explanation for the optical modulations is rotation combined with a non-uniform accretion of the metals that we have, in fact, detected in the UV spectrum of this WD.

In the case of WD\,J1857+4909, a massive $\approx 1$\,\msun\ WD, we can place an upper limit of $M_2 \sin i \lesssim 0.19$\,\msun\ on any companion based on the $2.2$\,h orbital period. This is only marginally consistent with a beaming explanation as the cause for the optical variations, as beaming would require $M_2 \sin i > 0.25$\,\msun\ \citep{Maoz_2015}.  \citet{Ostensen_2011} have suggested unresolved Zeeman splitting from a weak magnetic field as the cause for the line broadening in the optical spectrum of this WD. The atmospheric fit tends to overestimate the mass in the presence of unresolved Zeeman splitting, but the mass would still be relatively high. Reflection from a giant planet is not a viable solution because such a planet's orbit would be well within the WD's tidal disruption radius. A brown dwarf companion with a very high albedo could marginally have a low enough effective temperature to satisfy the $T_\textrm{eff}<1000$\,K limit from the SED, while producing the optical modulation via reflection of the WD light. Since no metals are detected, within our sensitivity limits, in the photosphere of this WD, there is no evidence for modulation that is associated with an uneven photospheric distribution of accreted metals, combined with rotation.

WD\,2359$-$434 has $M \approx 0.85$\,\msun, and RV measurements using the narrow H$\alpha$ NLTE absorption line core were made by \citet{Maxted_1999}. When folded over the observed period, these show no significant RV variations \citep{Gary_2013} down to $\approx 20$\,km\,s$^{-1}$, implying $M_2 \sin i \lesssim 45$\,M$_\textrm{Jup}$, i.e. the RV data do not rule out a cool ($<1000$\,K, from the SED constraints, above) brown dwarf companion. However, the variability in this cool ($T_\textrm{eff} \approx 8600$\,K) and hence likely convective-atmosphere WD could simply be the result of star spots combined with rotation (but see Section~\ref{sec:WD2359} for a possible conflict of this picture with the observed UV spectrum).

\section{Discussion and conclusions}
\label{sec:Discussion}

\begin{figure}
\centering
\includegraphics[width=\columnwidth]{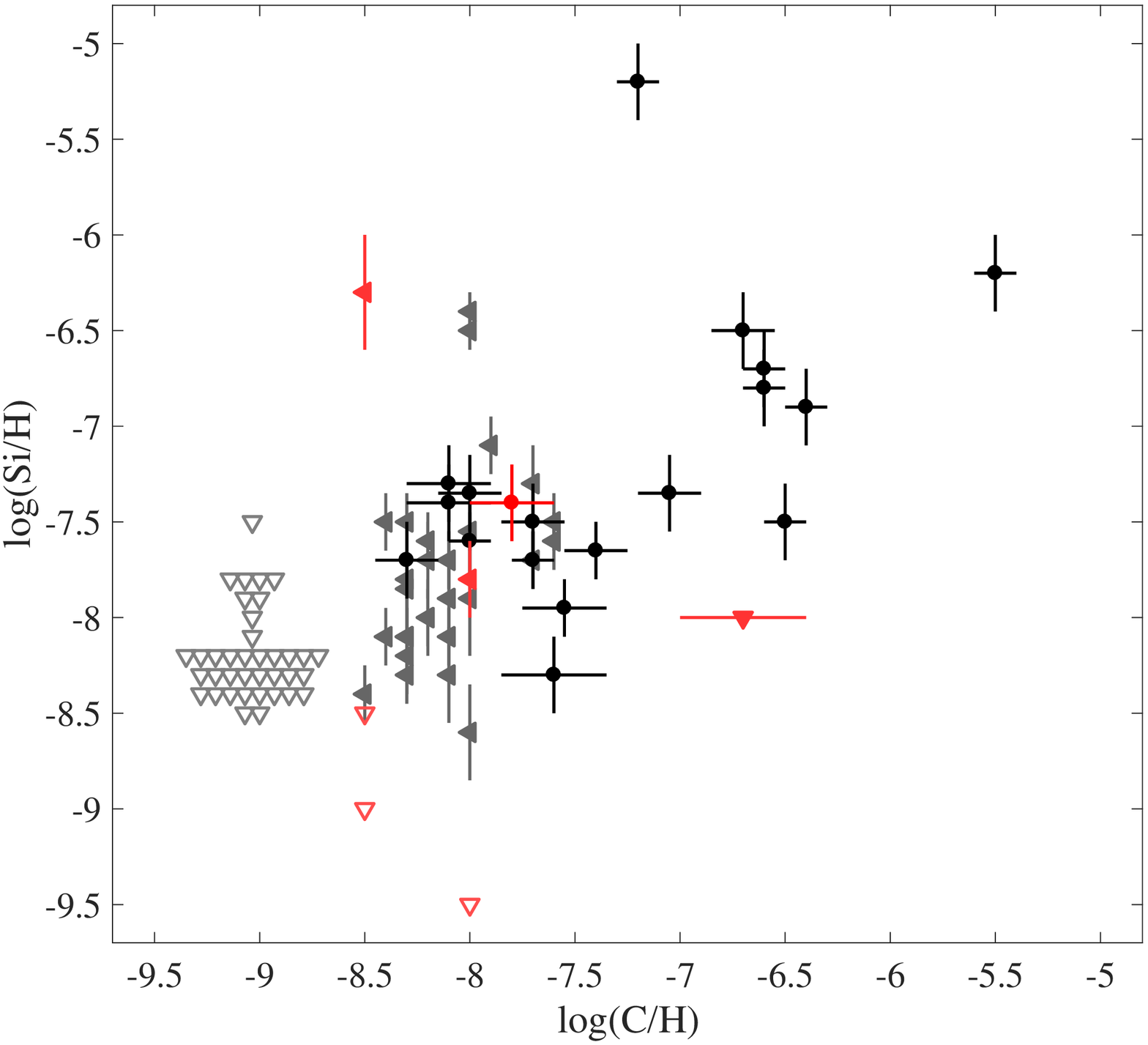}
\caption{Si and C abundances relative to H, in the \citet{Koester_2014} sample (grayscale symbols), and in our sample (red). Actual detections are represented by errorbars along the appropriate dimension, while upper limits are represented by triangle-shaped symbols.
For WD's with only upper limits on \abd{Si}, no upper limits were reported by \citet{Koester_2014} for \abd{C}, so we plot them at an arbitrary \abd{C} $ =-9$, with some small scatter added to avoid clutter in the figure. The upper limits in the \citet{Koester_2014} sample are at abundance values similar to those of the actual detections, suggesting that a larger fraction of the sample, perhaps approaching 100~per cent, might reveal photospheric metals at similar abundances, if observed at higher sensitivity.}
\label{fig:Abundances}
\end{figure}

We have obtained \textit{HST} UV spectra of seven WDs in order to test for a connection between debris accretion and periodic optical variability. The results of our experiment, and especially the presence of metal absorption lines in non-variable WDs, indicate that our null hypothesis, that optical variability and debris accretion are not directly linked, cannot be rejected. The cause of low-level periodic variability in a large fraction of WDs remains unclear. However, despite the results of our experiment, rotation combined with an inhomogeneous atmospheric metal distribution could still be the mechanism behind the variability in some or all cases. Fig. 8 of \citet{Koester_2014} shows that, in many WD with UV spectroscopy, the upper limits on Si abundance are comparable to the levels of the actual detections in other WDs. Metals at similar abundance levels could therefore be present in most or even all WDs. To illustrate this, we re-plot in Fig.~\ref{fig:Abundances} both the Si and C abundances, or upper limits thereof, for the \citet{Koester_2014} WD sample, as tabulated by them, and add the measurements and upper limits from the present work. In our sample, in the variable, but ``metal free'', WDs, the metal abundances could be non-zero but simply below the limits that our observations are sensitive to.

Looking back at the atmospheric parameters of the WDs in our sample (Table~\ref{tab:Results}), it appears that all of the non-variable WDs have `normal' parameters of $\log g \sim 8$ and a non-convective effective temperature, while each of the variable WDs is somewhat abnormal: WD\,J1855+4207 has a relatively low mass that might indicate binarity (or at least a binary past), WD\,J1857+4909 is massive with some indication of a moderate magnetic field, and WD\,2359$-$434 is magnetic and possibly convective. The determining factor linking debris accretion to optical variability could thus be the strength of a magnetic field such that accretion can be channelled inhomogeneously over the WD surface. Alternatively, magnetic fields might be behind the WD surface inhomogeneity of variable WDs, but without the intercession of photospheric metals. Subtle magnetic-field-sensitive radiative-transfer effects might exist, that when combined with rotation, could produce the low, $\sim 10^{-4}$, observed photometric modulation levels.

Our observations include two cases of photospherically polluted hot WDs that  show high-ionization lines from apparently circumstellar material that is perhaps infalling, or alternatively, in a layer not far above the photosphere, such that the lines have velocities similar to, or slightly blue-shifted compared to the gravitationally redshifted photospheric lines. The two WDs are similar in this aspect to several others recently reported (see Section~\ref{sec:WD1855}). The circumstellar material, if that is indeed its nature, may provide a new and important perspective on WD debris accretion.

More sensitive observations are required to resolve the puzzle of periodic optical variability. Deeper UV spectra could reveal up to what degree photometric metals are actually ubiquitous in WDs. Higher-resolution optical spectroscopy can set firmer constraints on, or reveal, binary companions inducing the optical modulation. Such high-resolution spectra could also illuminate the connection, if one exists, to magnetic field strength.

\section*{Acknowledgements}
We thank Gilles Fontaine for his help with the modelling of J1926+4219.
We thank our contact scientist, Cristina Oliveira, and our programme coordinator, Amber Armstrong, at the Space Telescope Science Institute (STScI), for their help in carrying out the observations. NH thanks Elaine Snyder from the COS Team Help Desk for her help with the \textsc{calcos} pipeline, Siyi Xu for her help with \textsc{Tlusty}, and Yossi Shvartzvald and Eran Ofek for their help with the \textit{Spitzer} data. The anonymous referee is thanked for valuable comments.
Based on observations made with the NASA/ESA \textit{Hubble Space Telescope}, obtained at the Space Telescope Science Institute, which is operated by the Association of Universities for Research in Astronomy, Inc., under NASA contract NAS 5-26555. These observations are associated with programmes 14082 (PI: Maoz), 11526 (PI: Green), and 14076 (PI: G\"{a}nsicke).
Support for programme 14082 was provided by NASA through a grant from the Space Telescope Science Institute, which is operated by the Association of Universities for Research in Astronomy, Inc., under NASA contract NAS 5-26555.
Based on observations obtained at the MMT Observatory, a joint facility of the Smithsonian Institution and the University of Arizona, and on observations obtained with the Apache Point Observatory 3.5\,m telescope, which is owned and operated by the Astrophysical Research Consortium.
This work was supported by Grant 648/12 by the Israel Science Foundation (ISF) and by Grant 1829/12 of the I-CORE programme of the PBC and the ISF.
The research leading to these results has received funding from the European Research Council under the European Union's Seventh Framework Programme (FP/2007-2013) / ERC Grant Agreement n. 320964 (WDTracer).
This research has made use of the VizieR catalogue access tool, CDS, Strasbourg, France.
This work is based in part on observations made with the \textit{Spitzer Space Telescope}, obtained from the NASA/IPAC Infrared Science Archive, both of which are operated by the Jet Propulsion Laboratory, California Institute of Technology under a contract with the National Aeronautics and Space Administration.
This work used the astronomy and astrophysics package for \textsc{Matlab} \citep{Ofek_2014}.




\bibliographystyle{mnras}
\bibliography{hstcospaper}




%
%


\bsp	
\label{lastpage}
\end{document}